\documentclass[lettersize,journal]{IEEEtran}
\onecolumn
\usepackage{amsmath}        
\usepackage{empheq}
\usepackage{mathtools}      
\usepackage{amssymb}        
\usepackage{mathrsfs}       
\usepackage{stmaryrd}       
\usepackage{bm}             
\usepackage{amsfonts}       
\usepackage{physics}		
\usepackage{cite}           
\usepackage{url}            
\usepackage{hyperref}       

\usepackage{enumitem}       

\usepackage{amsthm}         
\usepackage{thmtools}       
\usepackage{nicematrix}     
\usepackage{geometry}
\usepackage{spalign}
\usepackage{booktabs}   
\usepackage{subcaption}
\usepackage{threeparttable}
\usepackage{multirow}

\theoremstyle{plain}
\newtheorem{theorem}{Theorem}[section]
\newtheorem{lemma}[theorem]{Lemma}
\newtheorem{corollary}[theorem]{Corollary}
\newtheorem{proposition}[theorem]{Proposition}
\newtheorem{conjecture}{Conjecture}[section]

\theoremstyle{definition}

\newtheorem{claim}{Claim}[section]

\newcommand\setnd[1]{ \left\{ #1 \right\} }
\newcommand\floor[1]{\left\lfloor {#1} \right\rfloor}

\newcommand\card[1]{\left|{#1}\right|}

\newcommand\ZZ{\mathbb{Z}}
\newcommand\CC{\mathbb{C}}
\newcommand\RR{\mathbb{R}}

\newcommand{\defeq}{\triangleq}

\newcommand{\factorial}[1]{#1!}
\newcommand{\binomial}[2]{{#1 \choose #2}}

\newcommand{\EE}{\mathcal{E}}

\DeclareMathOperator{\wt}{\mathrm{wt}}

\makeatletter  
\let\oldtheequation\theequation
\renewcommand\tagform@[1]{\maketag@@@{\ignorespaces#1\unskip\@@italiccorr}}
\renewcommand\theequation{(\oldtheequation)}
\makeatother   

\title{Linear Programming Bounds on $k$-Uniform States}
\author{Yu Ning, Fei Shi, Tao Luo, and Xiande Zhang%

\thanks{The research of Yu Ning, Tao Luo and Xiande Zhang was
supported in part by the Innovation Program for Quantum Science and Technology
under Grant 2021ZD0302902, in part by NSFC under Grant 12171452 and
Grant 12231014, and in part by the National Key Research and Development
Programs of China under Grant 2023YFA1010200 and Grant 2020YFA0713100.
\emph{(Corresponding author: Xiande Zhang (e-mail: drzhangx@ustc.edu.cn)).}
}%

\thanks{Fei Shi acknowledges funding from National Natural Science Foundation
of China (NSFC) via Project No. 12347104 and No. 12305030, Guangdong Natural
Science Fund via Project 2023A1515012185, Hong Kong Research Grant Council
(RGC) via No. 27300823, N\_HKU718/23, and R6010-23, Guangdong Provincial
Quantum Science Strategic Initiative No. GDZX2303007, HKU Seed Fund for Basic
Research for New Staff via Project 2201100596.}%

\thanks{Yu Ning is with Hefei National Laboratory, Hefei 230088, China
(e-mail: sirning@mail.ustc.edu.cn)}%

\thanks{Fei Shi is with QICI Quantum Information and Computation Initiative,
	School of Computing and Data Science, The University of Hong Kong, Pokfulam
	Road, Hong Kong, China (e-mail: shifei@mail.ustc.edu.cn)}%

\thanks{Tao Luo and Xiande Zhang are with the School of Mathematical Sciences,
	University of Science and Technology of China, Hefei 230026, Anhui, China,
	and also with Hefei National Laboratory, University of Science and
	Technology of China, Hefei 230088, China
	(e-mail: luotao01@mail.ustc.edu.cn; drzhangx@ustc.edu.cn)}%
}

\begin{document}
\maketitle
\begin{abstract}
The existence of $k$-uniform states has been a widely studied problem due to 
their applications in several quantum information tasks and their close 
relation to combinatorial objects like Latin squares and orthogonal arrays. 
With the machinery of quantum enumerators and linear programming, we establish 
several improved non-existence results and bounds on $k$-uniform states.
\begin{itemize}
\item First, for any fixed $l\geq 1$ and  $q\geq 2$, we show that there exists 
a constant $c$ such that $(\floor{n/2}-l)$-uniform states in
$(\CC^q)^{\otimes n}$ do not exist when $n\geq cq^2+o(q^2)$. The constant $c$ 
equals $4$ when $l=1$ and $6$ when $l=2$, which generalizes Scott's 
bound (2004) for $l=0$. 
\item Second, when $n$ is sufficiently large, we show that there 
exists a constant $\theta<1/2$ for each $q \le 9$, such that $k$-uniform states 
in $(\CC^q)^{\otimes n}$ exist only when $k\leq \theta n$. In particular, this 
provides the first bound (to the best of our knowledge) of $k$ for 
$4\leq q\leq 9$ and confirms a conjecture posed by Shi \emph{et al}. 
(2023) when $q=5$ in a stronger form. 

  \item Finally, we improve the shadow bounds given by Shi \emph{et al}. (2023) 
  by a constant for $q = 3,4,5$ and small $n$. 
  When $q=4$, our results can update some bounds listed in the code tables 
  maintained by Grassl (2007--2024).
\end{itemize}

\end{abstract}	

\section{Introduction}
Mutipartite entanglement is an important resource for quantum information tasks. As a class of highly entangled
pure states in mutipartite systems,
$k$-uniform states have the property that  all $k$-party reductions are maximally mixed
\cite{Scott2004}. Due to their  applications in quantum secret sharing \cite{Helwig2012}
and quantum information masking \cite{Shi2021,Shi2022Heterogeneous}, the existence and construction problems of  $k$-uniform states have been widely studied. For example, the Greenberger–Horne–Zeilinger (GHZ) state is a $1$-uniform
state. General $k$-uniform states
 are equivalent to quantum
error-correcting codes (QECCs) with dimension one \cite{Scott2004}.
 Some
systematic constructions of $k$-uniform states are related to combinatorial
designs \cite{Goyeneche2014,Goyeneche2015,LiMaosheng2019,Shi2022Heterogeneous},
graph states \cite{Feng2017} and classical error-correcting codes
\cite{Feng2017,Raissi2020}. 

Besides the existence results, there are also many non-existence
results on $k$-uniform states. By Schmidt decomposition or the quantum Singleton bound
\cite{Ashikhmin1999,Grassl2022}, one can easily show that $k \le \floor{n/2}$
for $k$-uniform states in $(\CC^q)^{\otimes n}$.
In particular, $\floor{n/2}$-uniform states in $(\CC^q)^{\otimes n}$ are called
absolutely maximally entangled (AME) states \cite{Helwig2012}.

Scott \cite{Scott2004} showed that AME states in $(\CC^q)^{\otimes n}$ do not exist when
$n\gtrapprox 2q^2$, i.e., $n$ is large compared to $q$ \footnote{For functions $f$ and $g$, we denote $f\gtrapprox g$
if $|f|\geq g+o(g)$, and denote $f\gtrsim g$
if $|f|\geq g+c$ for some constant $c$.}. 
Huber \emph{et al}. \cite{Huber2018AME} further showed the non-existence
 of AME states for some small $n$ when $q = 3,4,5$. 
For qubit systems, i.e., $q=2$, the existence  of AME
states has been completely determined: AME states in $(\CC^2)^{\otimes n}$ exist if and only if $n =
2,3,5,6$ \cite{Rains1998Shadow,Scott2004,Huber2017AME,Huber2018AME}.
For larger $q$, many cases remain open, and we refer to the online table
of AME states maintained by Huber and Wyderka \cite{Huber:AMEtable}.
For general $k$-uniform
states, when $k\gtrsim n/3$, the non-existence of $k$-uniform
states in qubit systems was confirmed by Rains' shadow bounds \cite{Rains1999QuantumShadow}, which was
further improved by a constant in
\cite{Han2006Nonexistence,Han2009Nonexistence}.
Recently, Shi \emph{et al}. \cite{Shi2023} extended the shadow bounds to
qutrit systems, and showed the non-existence of $k$-uniform
states in qutrit systems when $k\gtrsim 3n/7$. For small parameters, we also refer to the online code tables
maintained by Grassl \cite{Grassl:codetables}.

The non-existence of $k$-uniform states by Rains \cite{Rains1998Shadow}, Huber
\emph{et al}. \cite{Huber2018AME} and
Shi \emph{et al}. \cite{Shi2023} is closely related to the quantum shadow
enumerator \cite{Rains1999QuantumShadow,Rains1998Shadow}.
Although this approach successfully yields some non-existence results of $k$-uniform states
for $q\leq 5$, there seems to be an obstacle for larger $q$.
For example, Huber \emph{et al}.  \cite{Huber2018AME} did not give the non-existence of any AME states
for $q \ge 6$. Similarly  for $q \ge 6$, Shi \emph{et al}.  \cite{Shi2023} did not give  the non-existence of any $k$-uniform states with $k \le \floor{n/2}$.

In this paper, we aim to establish  bounds on $k$-uniform states for larger
$q$.
The main approach is to set up a linear
program
based on the constraints  on the quantum weight enumerator and the unitary
enumerator of $k$-uniform states. Then we analyze the linear
program itself or its dual program to get the non-existence results. 
We also refine the method by the quantum shadow enumerator used in 
\cite{Rains1998Shadow,Shi2023}. 
Our contributions can be summarized below.
\begin{itemize}
 \item[(1)] We extend Scott's bound from AME states to $k$-uniform states with 
  $k=\floor{n/2}-1$ or $k=\floor{n/2}-2$. In particular, we show that 
  $(\floor{n/2}-1)$-uniform and $(\floor{n/2}-2)$-uniform states in
$(\CC^q)^{\otimes n}$ do not exist when $n\gtrapprox4q^2$ and 
$n\gtrapprox6q^2$, respectively.  

  \item[(2)] For any fixed $l\geq 3$, we show that $(\floor{n/2}-l)$-uniform states in
$(\CC^q)^{\otimes n}$ do not exist when $n\gtrapprox cq^2$, where $c$ is a constant only depending on $l$. We conjecture that  $c= 2l+2$. 

\item[(3)] For the case when $l$ is linear on $n$ and $n$ is sufficiently large, we show that there exists a constant $\theta<1/2$ for each $q\leq 9$, such that $k$-uniform states in
$(\CC^q)^{\otimes n}$ exist only when $k\leq \theta n$. Especially, when $q=4$, $\theta=0.479$, which provides the first bound of $k$ in this case; when $q=5$, $\theta=0.487$, which confirms a
conjecture in \cite[Conjecture 9(2)]{Shi2023} in a stronger form.
  
\item[(4)] Finally, we improve the shadow bounds for $k$-uniform states listed in \cite{Shi2023} by a constant for $q = 3,4,5$. Note that when $q=4$, we even improve some bounds listed in the code tables by Grassl
\cite{Grassl:codetables}.
\end{itemize}


This paper is organized as follows. \autoref{sec:preliminary} provides
preliminary knowledge. \autoref{sec:defect1} focuses on
the extension of Scott's bound to $(\floor{n/2}-l)$-uniform states for $l=1,2$.
\autoref{section:dual} is devoted to the dual linear program, where
the results for $l\geq 3$ and for $l$ linear on $n$
 are  derived. In \autoref{sec:extremal},
we improve the shadow bounds for $k$-uniform states in $(\CC^q)^{\otimes n}$
for $q = 3,4,5$. Finally, we conclude this paper in \autoref{sec:conclusion}.

\section{Preliminary}
\label{sec:preliminary}
In this section, we introduce necessary preliminary knowledge, including known 
bounds of $k$-uniform states, various kinds of quantum enumerators and a linear
program for the existence of $k$-uniform states.

\subsection{$k$-Uniform States}
A $k$-\emph{uniform} state $\ket{\psi}$ in $(\CC^q)^{\otimes n}$ \cite{Scott2004}, is a pure
state whose $k$-party reductions are all maximally mixed,
that is, the partial trace $\tr_T(\rho) = \frac{I}{q^k}$ for all $T \subset [n] \defeq
\setnd{1,2,\dots,n}$ with $\card{T} = n-k$, where
$\rho = \ketbra{\psi}$, and $I$ is the identity operator over the corresponding
space $(\CC^q)^{\otimes k}$. It is known that 
$k \le \floor{n/2}$ when a $k$-uniform state in $(\CC^q)^{\otimes n}$
exists \cite{Ashikhmin1999,Grassl2022}.
In particular, $\floor{n/2}$-uniform states in $(\CC^q)^{\otimes n}$ are called \emph{absolutely
maximally entangled (AME)} states.
Scott \cite{Scott2004} proved the following non-existence of AME states when
$n$ is big compared to the local dimension $q$.
\begin{theorem}[Scott's bound \cite{Scott2004}]
	\label{theorem:Scott}
	AME states in $(\CC^q)^{\otimes n}$ do not exist, if
	$$
	n > \begin{cases}
		2(q^2-1), & n \text{ even}, \\
		2q(q+1)-1, & n \text{ odd}.
	\end{cases}
	$$
\end{theorem}
Huber \emph{et al}. \cite{Huber2018AME} further presented the following
non-existence of AME states for small $n$ by the shadow inequality.
\begin{theorem}[\cite{Huber2018AME}]
	\label{theorem:Huber}
	AME states in $(\CC^q)^{\otimes n}$ do not exist when
	\begin{itemize}
		\item $q = 3$ and $n = 8, 12, 13, 14, 16, 17, 19, 21, 23$;
		\item $q = 4$ and $n = 12, 16, 20, 24, 25, 26, 28, 29, 30, 33, 37, 39$;
		\item $q = 5$ and $n = 28, 32, 36, 40, 44, 48$.
	\end{itemize}
\end{theorem}
The non-existence of AME states  implies that $k\le \floor{n/2}-1$ when a
$k$-uniform state exists. When $q = 2,3$, the following shadow bounds
provided with more constraints on $k$. Here, $[a,b]$ stands for
$\setnd{a,a+1,\dots,b}$ for any integers $a \le b$.

\begin{theorem}
	\label{theorem:Rains}
For a $k$-uniform state in $(\CC^q)^{\otimes n}$, the followings hold.
\begin{itemize}
  \item[(1)] (Rains' shadow bound \cite{Rains1998Shadow}) When $q=2$, $k \le
	\begin{cases}
		2m + 1, & n \in [6m,6m+4], \\
		2m + 2, & n = 6m + 5,
	\end{cases}
	$
	for any $m \ge 0$.
  \item[(2)] (Shi \emph{et al}. \cite{Shi2023}) When $q=3$,  $k \le \begin{cases}
		6m-1, & n \in [14m-4,14m-1], \\
		6m+1, & n \in [14m,14m+4], \\
		6m+3, & n \in [14m+5,14m+9],
	\end{cases}
	$
	where $m \ge 1$ and $n \notin \setnd{23,37,51}$.
\end{itemize}
\end{theorem}
%
For $q = 2$, Rains' shadow bound in \autoref{theorem:Rains} (1) can be further
improved by a constant \cite{Han2006Nonexistence,Han2009Nonexistence}.
For $q = 4,5$, Shi \emph{et al}. \cite{Shi2023} raised the following
conjectures, and showed the correctness for small $n$. See \cite[Table II and 
Table III]{Shi2023}.

\begin{conjecture}[\cite{Shi2023}]
	\label{conj:q_4}For a $k$-uniform state in $(\CC^q)^{\otimes n}$, the followings hold.
\begin{itemize}
  \item[(1)] When $q=4$, $k \le
	\begin{cases}
		8m - 5, & n \in [17m-12,17m-9], \\
		8m - 3, & n \in [17m-8,17m-5], \\
		8m - 1, & n \in [17m-4,17m-1], \\
		8m + 1, & n \in [17m,17m+4],
	\end{cases}$
	where $m \ge 2$, except for $n = 38$.
  \item[(2)] When $q=5$, $
	k \le 2m -1$ for $ n \in [4m,4m+3],
	$
	where $m \ge 45$.
\end{itemize}
\end{conjecture}


We refer all the bounds of $k$ in \autoref{theorem:Rains} and 
\autoref{conj:q_4} as the shadow bounds.  Note that the shadow bounds are 
stronger than Scott's bound for specific $q\leq 5$. However, Scott's bound 
provides bounds of $k$ for larger $q$. 

In this paper, combining the advantages of both the shadow bounds and Scott's bound, we show the non-existence of $(\floor{n/2}-l)$-uniform states for large $l$ and for large $q$. For
convenience, we call $(\floor{n/2}-l)$-uniform states as AME states of \emph{defect} $l$. We further improve the shadow bounds for $q = 3,4,5$  by a constant. Our main tools are various kinds of quantum enumerators
\cite{Shor1997,Rains1998QuantumWeight,Rains1999QuantumShadow,Calderbank1998},
which will be recalled in the next subsection.


\subsection{Quantum Enumerators}

We follow the notations in
\cite{Huber2018AME,Rains1998QuantumWeight,Rains1999QuantumShadow}, and
focus on the enumerators associated with a pure state. Let $\setnd{e_j}_{j \in
\ZZ_{q^2}}$, with $e_0 = I$, be an orthogonal list of error basis of linear
operators acting on $\CC^q$, with respect to the trace inner product, i.e.,
$\tr(e_i^\dagger e_j) = \delta_{i,j}q$. For the qubit system, the Pauli
operators serve as such a basis. For $q \ge 3$, the existence of such an error
basis is assured by the Weyl operators \cite{Bertlmann2008Bloch}.
Let $\EE$ be the set of $n$-fold tensor products of the error basis, i.e.,
an element $E \in \EE$ is of the form
$$
E = e_{j_1} \otimes e_{j_2} \otimes \dots \otimes e_{j_n},
$$
where $j_1,\dots,j_n \in \ZZ_{q^2}$. Naturally, $\EE$ serves as an orthogonal
list of error basis for the space
$(\CC^q)^{\otimes n}$. For $E \in \EE$,
let $\wt(E)$ denote the weight of $E$, that is
the number of tensor factors in $E$ not equal to the identity.

Let $\ket{\psi} \in (\CC^q)^{\otimes n}$ be a pure state, and
$\rho = \ketbra{\psi}$ be its density matrix. The \emph{quantum weight
distribution}
associated with $\ket{\psi}$ is defined as
\begin{equation}
	A_i \defeq \sum_{\wt(E) = i} \tr(E\rho)\tr(E^\dagger \rho), \quad
	B_i \defeq \sum_{\wt(E) = i} \tr(E\rho E^\dagger \rho), \quad i \in [0,n],
\end{equation}
where the index of summand ranges over all elements in $\mathcal{E}$ of
weight $i$.
The associated \emph{quantum weight enumerators} are defined as
\begin{equation}
	A(x,y) \defeq \sum_{i=0}^nA_ix^{n-i}y^i, \quad
	B(x,y) \defeq \sum_{i=0}^nB_ix^{n-i}y^i.
\end{equation}
Then $A(x,y)$ and $B(x,y)$ satisfy the quantum MacWilliams transformation
\cite{Shor1997,Huber2018AME}
\begin{equation}
	B(x,y) = A\qty(\frac{x+(q^2-1)y}{q},\frac{x-y}{q}).
\end{equation}
Also, as $\ket{\psi}$ is a pure state, we have that $A_0 = 1$ and $A_i=B_i \ge
0$ for all $i \in [0,n]$, and thus $A(x,y) = B(x,y)$ \cite{Shi2023}.
Moreover, if $\ket{\psi}$ is $k$-uniform, we have that
$A_1 = A_2 = \dots = A_k = 0$ \cite{Shi2023}. If $\ket{\psi}$ is a
stabilizer state, $A(x,y)$
has the combinatorial interpretation as the weight enumerator of the
corresponding stabilizer group regarded as an additive code
\cite{Rains1999QuantumShadow}.

Next, we recall the unitary enumerator associated with $\ket{\psi}$
\cite{Rains1998QuantumWeight,Huber2018AME}.
For $T \subset [n]$, we write $\rho_T = 
\tr_{T^\complement}(\rho)$, where $T^\complement$ is the complement of 
$T$ in $[n]$, and define
\begin{equation}
	A_T^\prime = \tr(\rho_T^2) \text{ and }
	A_i^\prime = \sum_{\card{T} = i} A_{T}^\prime, \quad i \in [0,n], 
\end{equation}
where the index of summand ranges over all subsets of $[n]$ of size $i$.
The \emph{unitary enumerator} associated with $\ket{\psi}$ is defined as
\begin{equation}
	A^\prime(x,y) = \sum_{i=0}^n A_i^\prime x^{n-i}y^i.
\end{equation}
The unitary enumerator and the quantum weight enumerator are related as
\cite{Huber2018AME}
\begin{equation}
	A(x,y) = A^\prime(x-y,qy), \quad A^\prime(x,y) =
	A\qty(x+\frac{y}{q},\frac{y}{q}).
\end{equation}
As $\ket{\psi}$ is a pure state, by Schmidt decomposition, we have that
\cite{Huber2020quantum}
\begin{equation}
A_T^\prime = A_{T^\complement}^\prime, \quad A_i^\prime = A_{n-i}^\prime,
\quad A^\prime(x,y) = A^\prime(y,x),
\end{equation}
for arbitrary $T \subset [n]$ and $i \in [0,n]$.
Also by Schmidt decomposition, we have that \cite{Huber2020quantum}
\begin{equation}
	A_T^\prime \ge \frac{1}{q^{\min(\card{T},n-\card{T})}}, \quad
	A_i^\prime \ge \frac{1}{q^{\min(i,n-i)}}\binomial{n}{i},
\end{equation}
where $T \subset [n]$ and $i \in [0,n]$ are arbitrary.
Moreover, if $\ket{\psi}$ is $k$-uniform, we have that \cite{Huber2020quantum}
\begin{equation}
	A_T^\prime = \frac{1}{q^{\card{T}}}, \quad
	A_i^\prime = \frac{1}{q^{i}}\binomial{n}{i},
\end{equation}
where $T \subset [n]$ is of size at most $k \le \floor{n/2}$ and $i \in [0,k]$.

Finally, we recall the shadow enumerator \cite{Rains1999QuantumShadow,Huber2018AME}.
Let
\begin{equation}
	S_i = \sum_{\card{R} = i}\sum_{T \subset [n]} (-1)^{\card{T \cap
	R^\complement}}A_T^\prime,
\end{equation}
where the summand $R$ ranges over all subsets of $[n]$ of size $i$.
The \emph{shadow enumerator} associated with $\ket{\psi}$ is defined as
\begin{equation}
	S(x,y) = \sum_{i=0}^nS_ix^{n-i}y^i.
\end{equation}
The shadow enumerator is related to the quantum weight enumerator and the unitary enumerator as in \cite{Huber2018AME}
\begin{equation}
	S(x,y) = A\qty(\frac{(q-1)x+(q+1)y}{q},\frac{y-x}{q}), \quad S(x,y) =
	A^\prime(x+y,y-x).
\end{equation}
As $\ket{\psi}$ is a pure state, $S_j \ge 0$ for all $j \in [0,n]$
and $S_{n-j} = 0$ for odd $j \in [0,n]$ \cite{Huber2018AME}.
Also, if $\ket{\psi}$ is a stabilizer state, $S(x,y)$ has the
combinatorial interpretation as the weight enumerator of the
shadow code of the stabilizer group regarded as an additive code
\cite{Rains1999QuantumShadow}.

\subsection{A Linear Program}
The transformations among the enumerators of $\ket{\psi}$ and
the $k$-uniform condition will impose linear constraints on the
coefficients of enumerators of $\ket{\psi}$.
The linear program concerning the quantum weight enumerator and the shadow
enumerator has been considered by many authors
\cite{Calderbank1998,Rains1998QuantumWeight,Rains1999QuantumShadow,
	Rains1998Shadow,Shi2023,Huber2018AME}.
In this paper, we will focus on the linear program related to the
quantum weight enumerator and the unitary enumerator, which we formulate
precisely in the following.

Let $\ket{\psi} \in (\CC^q)^{\otimes n}$ be a $k$-uniform state,
and $A(x,y) = \sum_{i=0}^n A_ix^{n-i}y^i$, $A^\prime(x,y) =
\sum_{i=0}^nA_i^\prime x^{n-i}y^i$ be the
weight enumerator and the unitary enumerator of $\ket{\psi}$, respectively. Since
\begin{equation}
	\sum_{i=0}^nA_i^\prime x^{n-i}y^i = A^\prime(x,y) = A\qty(x+\frac{y}{q},\frac{y}{q})
	= \sum_{j=0}^n A_j \qty(x+\frac{y}{q})^{n-j} \qty(\frac{y}{q})^j,
\end{equation}
by comparing the coefficients on both sides, we have
\begin{equation}
	q^iA_i^\prime = \sum_{j = 0}^{i} A_j \binom{n-j}{i-j}, \qquad i \in [0,n].
\end{equation}
Then, we have the following linear constraints on enumerators for a $k$-uniform
state \cite{Huber2018AME}:
\begin{empheq}[left=\empheqlbrace]{align}
&A_0 = 1; \quad A_1=A_2=\dots=A_k = 0; \quad A_i \ge 0, \,\forall i
\in [k+1,n]; \label{eqn:unitary_constraints_1} \\
&A_i^\prime = \frac{1}{q^i}\binom{n}{i}, \,\forall i \in [0,k];\quad
A_i^\prime \ge \frac{1}{q^i}\binom{n}{i}, \,\forall i\in
[k+1,\floor{n/2}]; \quad
A_i^\prime = A_{n-i}^\prime, \,\forall i \in [0,n]; \label{eqn:unitary_constraints_2} \\
&q^iA_i^\prime = \sum_{j = 0}^{i} A_j \binom{n-j}{i-j}, \,\forall i \in
[0,n]. \label{eqn:unitary_constraints_3}	
\end{empheq}
If we can show that the constraints of
Eqs. \ref{eqn:unitary_constraints_1}--\ref{eqn:unitary_constraints_3}
do not hold, then $k$-uniform states in $(\CC^q)^{\otimes n}$ do
not exist.

\section{Scott-Type Bounds for AME States of Defect $1$ and $2$}
\label{sec:defect1}
In this section,
we aim to extend Scott's bound for AME states of defect $l\leq 2$, that is, 
show the non-existence of $(\floor{n/2}-1)$-uniform states and 
$(\floor{n/2}-2)$-uniform states in $(\CC^q)^{\otimes n}$ for large $n$.
The main result of this section is the following theorem.
\begin{theorem}[Scott-type bounds for AME states of defect $1$ and $2$]
	\label{theorem:defect1}
	Let $q \ge 2$.
\begin{itemize}
  \item[(1)]AME states of defect $1$
	in $(\CC^q)^{\otimes n}$ do not exist if
	$
	n > \begin{cases}
		4q^2, & n \text{ even}, \\
		4q^2 + 4q + 1, & n \text{ odd}.
	\end{cases}
	$
  \item[(2)] AME states of defect $2$ in $(\CC^q)^{\otimes n}$ do not
	exist if
	$
	n > \begin{cases}
		6q^2 + 2, & n \text{ even},	 \\
		6q^2 + 6q + 3, & n \text{ odd}.
	\end{cases}
	$
\end{itemize}

\end{theorem}
For $q=4,5,6,7$ and $l=1,2$, the lower bounds of $n$ in
\autoref{theorem:defect1} are listed in \autoref{tab:defect12}.
For $q = 5$, \autoref{theorem:defect1} improves some of the results
listed in \cite[Table III]{Shi2023}, which we compare in
\autoref{tab:small_defect_compare}.

\setcounter{table}{0}
\begin{table}
\centering
\caption{Non-Existence of AME States of Defect $l \in \setnd{1,2}$
	in $(\CC^q)^{\otimes n}$}
\label{tab:defect12}
	\begin{threeparttable}
		\begin{subtable}{.25\textwidth}
			\begin{tabular}{|l|l|l|}
				\toprule
				\multicolumn{3}{|c|}{$q = 4$} \\
				\midrule
				$l$ & $n$ even & $n$ odd  \\
				\midrule
				$1$ & $n \ge 66$ & $n \ge 83$ \\
				$2$ & $n \ge 100$ & $n \ge 125$ \\
				\bottomrule
			\end{tabular}
		\end{subtable}
		\begin{subtable}{.25\textwidth}
			\begin{tabular}{|c|c|c|}
				\toprule
				\multicolumn{3}{|c|}{$q = 5$} \\
				\midrule
				$l$ & $n$ even & $n$ odd  \\
				\midrule
				$1$ & $n \ge 102$ & $n \ge 123$ \\
				$2$ & $n \ge 154$ & $n \ge 185$ \\
				\bottomrule
			\end{tabular}
		\end{subtable}
		\begin{subtable}{.25\textwidth}
			\begin{tabular}{|c|c|c|}
				\toprule
				\multicolumn{3}{|c|}{$q = 6$} \\
				\midrule
				$l$ & $n$ even & $n$ odd  \\
				\midrule
				$1$ & $n \ge 146$ & $n \ge 171$ \\
				$2$ & $n \ge 220$ & $n \ge 257$ \\
				\bottomrule
			\end{tabular}
		\end{subtable}
		\begin{subtable}{.25\textwidth}
			\begin{tabular}{|c|c|c|}
				\toprule
				\multicolumn{3}{|c|}{$q = 7$} \\
				\midrule
				$l$ & $n$ even & $n$ odd  \\
				\midrule
				$1$ & $n \ge 198$ & $n \ge 227$ \\
				$2$ & $n \ge 298$ & $n \ge 341$ \\
				\bottomrule
			\end{tabular}
		\end{subtable}
		\end{threeparttable}
\end{table}
\setcounter{table}{1} 
\begin{table}
	\centering
	\begin{threeparttable}
			\caption{Improvement on Bounds of $k$-Uniform States in 
			$(\CC^5)^{\otimes
				n}$}
		\label{tab:small_defect_compare}
		\begin{tabular}{|c|ccccccccccccccc|}
			\toprule
			$n$ & $180$ & $182$ & $185$ & $187$ & $224$ & $226$ & $228$ & $261$
			& $263$ & $265$ & $267$ & $269$ & $271$ & $273$ & $275$\\
			\midrule
			$k \le $   & $87$ & $88$ & $89$ & $90$ & $108$ & $109$ & $110$ &
			$126$ & $127$ & $128$ & $129$ & $130$ & $131$ & $132$ & $133$\\
			\midrule
			$k \le $   & $89$ & $89$ & $91$ & $91$ & $111$ & $111$ & $111$ &
			$129$ & $129$ & $131$ & $131$ & $133$ & $133$ & $135$ & $135$\\
			\bottomrule
		\end{tabular}
	\begin{tablenotes}
		\footnotesize
		\item The upper bounds in the second row are from
		\autoref{theorem:defect1} (2) when $n \le 187$ and from \autoref{theorem:dual_unitary_supp_2} when $n \ge 224$.
		Also check \autoref{tab:defect12} and \autoref{tab:ame_small_defects}
		for comparision. Upper bounds in the last row are from
		\cite[Table III]{Shi2023}.
	\end{tablenotes}
	\end{threeparttable}
\end{table}

The rest of this section will be devoted to the proof of
\autoref{theorem:defect1}. The key point is to find when the constraints in
Eqs. \ref{eqn:unitary_constraints_1}--\ref{eqn:unitary_constraints_3}
do not hold.
We begin with the easy case, \autoref{theorem:defect1} (1).

\subsection{A Proof of \autoref{theorem:defect1} (1)}
Let $k = \floor{n/2}-1$, and assume that $\ket{\psi}$ is $k$-uniform.
Plugging the values of \autoref{eqn:unitary_constraints_1} and
\autoref{eqn:unitary_constraints_2} into \autoref{eqn:unitary_constraints_3}
for $i = k+2$ and $k+3$, we have
\begin{equation}
	\label{eqn:k2k3}
	\begin{aligned}
		q^{k+2}A_{k+2}^\prime&=\binom{n}{k+2} + A_{k+1}(n-(k+1)) + A_{k+2}, \\
		q^{k+3}A_{k+3}^\prime&=\binom{n}{k+3} + A_{k+1}\binom{n-(k+1)}{2} +
		A_{k+2}(n-(k+2)) + A_{k+3} \\
		&=q^{2(k+3)-n}\binom{n}{k+3}.
	\end{aligned}
\end{equation}
In the following, we will split into two cases where $n$ is even or odd.

When $n$ is even, that is, $n = 2k+2$, we have $A_{k+2}^\prime = A_{k}^\prime =
\frac{1}{q^{k}}\binom{n}{k+2}$ by \autoref{eqn:unitary_constraints_2}.
Solving $A_{k+2}$ and $A_{k+3}$ in
\autoref{eqn:k2k3},
we have that
\begin{equation}
\begin{aligned}
A_{k+2} &= \qty(q^{2}-1)\binom{2k+2}{k+2} - A_{k+1}(k+1), \\
A_{k+3} &= (q^4-1)\binom{2k+2}{k+3} - \binom{k+1}{2}A_{k+1} - kA_{k+2} \\
&= \qty(q^{4}-1)\binom{2k+2}{k+3} + A_{k+1}\binom{k+1}{2} -
k\qty(q^{2}-1)\binom{2k+2}{k+2}.
\end{aligned}
\end{equation}
With the constraints that $A_{k+2} \ge 0$ and $A_{k+3} \ge 0$, we have that
\begin{equation}
	\label{eqn:conclusion}
	\frac{q^2-1}{k+1}\binom{2k+2}{k+2} \ge A_{k+1} \ge
	\frac{1}{\binom{k+1}{2}}\qty(k(q^2-1)\binom{2k+2}{k+2}-(q^4-1)\binom{2k+2}{k+3}).
\end{equation}
So if \autoref{eqn:conclusion} is violated, i.e.,
\begin{equation}
	\label{eqn:condition}
	\frac{q^2-1}{k+1}\binom{k+1}{2}\binom{2k+2}{k+2} <
	k(q^2-1)\binom{2k+2}{k+2} - (q^4-1)\binom{2k+2}{k+3},
\end{equation}
then AME states of defect $1$ in $(\CC^q)^{\otimes n}$ do not exist.
Since \autoref{eqn:condition} holds when $k > 2q^2 - 1$, that is
$n=2k+2> 4q^2$, \autoref{theorem:defect1} (1)
follows for even $n$.

%

When $n$ is odd, that is, $n = 2k+3$, we still have that 
$A_{k+3}^\prime = A_{k}^\prime = \frac{1}{q^k}\binom{n}{k}$. 
But $A_{k+2}^\prime = A_{k+1}^\prime$ is unknown to us, which makes the
analysis a little bit more complicated.
We need one more equation for $i = k+1$ in \autoref{eqn:unitary_constraints_3},
i.e.,
\begin{equation}
	\label{eqn:k1}
	q^{k+1}A_{k+1}^\prime = \binom{n}{k+1} + A_{k+1}.
\end{equation}
Combining \autoref{eqn:k1} and \autoref{eqn:k2k3}, we have that
\begin{equation}
	\label{eqn:k1k2k3}
	\begin{aligned}
		q^{k+1}A_{k+2}^\prime = q^{k+1}A_{k+1}^\prime &= \binom{2k+3}{k+1} +
		A_{k+1}, \\
		q^{k+2}A_{k+2}^\prime &= \binom{2k+3}{k+2} + A_{k+1}(k+2) + A_{k+2}, \\
		q^3\binom{2k+3}{k+3} = q^{k+3}A_{k+3}^\prime &=
		\binom{2k+3}{k+3} + A_{k+1}\binom{k+2}{2} + A_{k+2}(k+1) + A_{k+3}.
	\end{aligned}
\end{equation}
Solving $A_{k+2},A_{k+3}$ out, we have that
\begin{equation}
	\label{eqn:k1k2k3_solution}
	\begin{aligned}
		A_{k+2} &= q^{k+2}A_{k+2}^\prime - \binom{2k+3}{k+1} - A_{k+1}(k+2) \\
		&= (q-1)\binom{2k+3}{k+1} - (k+2-q)A_{k+1}, \\
		A_{k+3} &= (q^3-1)\binom{2k+3}{k+3} - A_{k+1}\binom{k+2}{2} -
		A_{k+2}(k+1) \\
		&=(q^3-1)\binom{2k+3}{k+3} -
		(k+1)(q-1)\binom{2k+3}{k+1}+(k+1)\qty(\frac{k+2}{2}-q)A_{k+1}.
	\end{aligned}
\end{equation}
As $A_{k+1},A_{k+2},A_{k+3}$ are all non-negative, we have that
\begin{equation}\label{eq:23}
	\begin{aligned}
		(q-1)\binom{2k+3}{k+1} &\ge (k+2-q)A_{k+1}
		\ge (k+2-2q)A_{k+1} \\
		&\ge 2(q-1)\binom{2k+3}{k+1} -
		\frac{2(q^3-1)}{k+1}\binom{2k+3}{k+3}.
	\end{aligned}
\end{equation}
However, when $k > 2q^2 + 2q - 1$,
\begin{equation}
	(q-1)\binom{2k+3}{k+1}-\frac{2(q^3-1)}{k+1}\binom{2k+3}{k+3}>0,
\end{equation}
that is, \autoref{eq:23} does not hold. So when $n=2k+3> 4q^2 + 4q + 1$, AME states of defect $1$ in
$(\CC^q)^{\otimes n}$
do not exist, which proves \autoref{theorem:defect1} (1) for odd $n$.
\subsection{A Proof of \autoref{theorem:defect1} (2)}

Assume that $\ket{\psi}$ is $k$-uniform with $k=\floor{n/2}-2$. The approach is similar to that for  \autoref{theorem:defect1} (1). But we need to consider more relations in
\autoref{eqn:unitary_constraints_3}. For $i = k+1,k+2,k+3,k+4$,
the relations of \autoref{eqn:unitary_constraints_3} can
be written in the following matrix eqaution
\begin{equation}
\label{eqn:k1k2k3k4_matrix}
\spalignmat{
\binom{n-(k+1)}{0};
\binom{n-(k+1)}{1},\binom{n-(k+2)}{0};
\binom{n-(k+1)}{2},\binom{n-(k+2)}{1},\binom{n-(k+3)}{0};
\binom{n-(k+1)}{3},\binom{n-(k+2)}{2},\binom{n-(k+3)}{1},\binom{n-(k+4)}{0};}
\spalignmat{
		A_{k+1};A_{k+2};A_{k+3};A_{k+4}
} = \spalignmat{
		q^{k+1}A_{k+1}^\prime;
		q^{k+2}A_{k+2}^\prime;
		q^{k+3}A_{k+3}^\prime;
		q^{k+4}A_{k+4}^\prime;
	}-
	\spalignmat{
		\binom{n}{k+1};
		\binom{n}{k+2};
		\binom{n}{k+3};
		\binom{n}{k+4};
	}.
\end{equation}
Let $M$ denote the $4 \times 4$ matrix on the left hand side of \autoref{eqn:k1k2k3k4_matrix}.
We note that $M$ has inverse
\begin{equation}
	\label{eqn:M_inverse}
M^{-1} = \spalignmat[r]{
		\binom{n-(k+1)}{0};
		-\binom{n-(k+1)}{1},\binom{n-(k+2)}{0};
		\binom{n-(k+1)}{2},-\binom{n-(k+2)}{1},\binom{n-(k+3)}{0};
		-\binom{n-(k+1)}{3},\binom{n-(k+2)}{2},-\binom{n-(k+3)}{1},\binom{n-(k+4)}{0}
	}.
\end{equation}
Let $\mathbf{v}_i$ denote the $i$th column of $M^{-1}$, $i\in [1,4]$.

\vspace{0.2cm}

\textbf{When $n$ is even, $n = 2k+4$.} In this case, $A_{k+4}^\prime = A_k^\prime = \frac{1}{q^k}\binom{n}{k}$
is known to us, but
$A_{k+1},A_{k+2},A_{k+3},A_{k+4},A_{k+1}^\prime=A_{k+3}^\prime,A_{k+2}^\prime$
are unknown.
We rewrite \autoref{eqn:k1k2k3k4_matrix} as a system of linear equations in
terms of all these unknowns as
\begin{equation}
	\label{eqn:linear_equation}
	\begin{pNiceArray}{cccc|ccc}
		\Block{4-4}<\Large>{M} &&&&-q^{k+1} & &\\
		&&&&& -q^{k+2}&\\
		&&&&& &-q^{k+3}\\
		&&&&0  &0 &0\\
		\hline
		0 & 0 & 0 & 0          &1       &0&-1
	\end{pNiceArray}
	\spalignmat{
		A_{k+1};A_{k+2};A_{k+3};A_{k+4};A_{k+1}^\prime;A_{k+2}^\prime;A_{k+3}^\prime
	} =
	\spalignmat{
		-\binom{n}{k+1};-\binom{n}{k+2};-\binom{n}{k+3};(q^4-1)\binom{n}{k+4};0
	}.
\end{equation}
Let $L$ denote the $5 \times 7$ coefficient matrix in
\autoref{eqn:linear_equation}.
As $M$ is invertible, $\rank{L} = 5$. Thus, the solutions to
\autoref{eqn:linear_equation} constitute a coset of a
subspace of dimension $2$, which we find in the following. Define
\begin{equation}
	\mathbf{y}_1 = \spalignmat{q^{k+1}\mathbf{v}_1;1;0;0}, \quad
	\mathbf{y}_2 = \spalignmat{q^{k+2}\mathbf{v}_2;0;1;0}, \qand
	\mathbf{y}_3 = \spalignmat{q^{k+3}\mathbf{v}_3;0;0;1},
\end{equation} which are column vectors with seven coordinates.
By the definition of $\mathbf{v}_i$, we know that
\begin{equation}
	L(\mathbf{y}_1+\mathbf{y}_3) = 0, \quad L\mathbf{y}_2 = 0.
\end{equation}
That is, the null space of $L$ is spanned by $\mathbf{y}_1+\mathbf{y}_3$ and
$\mathbf{y}_2$ as a basis.
For a particular solution to \autoref{eqn:linear_equation},  say $\mathbf{y}_p =
(a_{k+1},a_{k+2},a_{k+3},a_{k+4},
a_{k+1}^\prime,a_{k+2}^\prime,a_{k+3}^\prime)^\intercal$, it is also a solution to \autoref{eqn:k1k2k3k4_matrix}.
We take the corresponding enumerators of an AME state (even though AME states
do not exist).
Define $a_{k+1}^\prime = a_{k+3}^\prime =
\frac{1}{q^{k+1}}\binom{n}{k+1}$ and
$a_{k+2}^\prime = \frac{1}{q^{k+2}}\binom{n}{k+2}$ by
\autoref{eqn:unitary_constraints_2}.
Substituting these values together with $A_{k+4}^\prime = \frac{1}{q^k}\binom{n}{k}$ 
into \autoref{eqn:k1k2k3k4_matrix}, we get
$a_{k+1}=a_{k+2}=0$ and
\begin{equation}\label{eqak34}
	a_{k+3} = (q^2-1)\binom{n}{k+3}, \quad a_{k+4} = (q^4-1)\binom{n}{k+4} -
	(n-(k+3))(q^2-1)\binom{n}{k+3}.
\end{equation}
It is clear that $\mathbf{y}_p$ defined above is a particular solution to \autoref{eqn:linear_equation}.
So any solution to \autoref{eqn:linear_equation} is of the form
\begin{equation}
	\label{eqn:y_even}
	\mathbf{y} = t_1(\mathbf{y}_1+\mathbf{y}_3) + t_2\mathbf{y}_2 + \mathbf{y}_p, \quad t_1,t_2 \in \RR.
\end{equation}
Since the unknowns in \autoref{eqn:linear_equation}  are enumerators of a $k$-uniform state,
there should be a choice of $t_1,t_2$ such that
each entry of $\mathbf{y}$ is non-negative.

\begin{claim}\label{cl:l2even}If $k > 3q^2-1$, there must be a negative entry in $\mathbf{y}$ in \autoref{eqn:y_even} for any choices of $t_1,t_2 \in \RR.$
\end{claim}
By \autoref{cl:l2even}, when $n=2k+4>6q^2+2$, $k$-uniform states in  
$(\CC^q)^{\otimes n}$ do not exist, that is, \autoref{theorem:defect1} (2) 
follows for even $n$.

\begin{proof}[Proof of \autoref{cl:l2even}]
The proof is by contradiction and we suppose that there is a choice of 
$t_1,t_2 \in \RR$ such that each entry of $\mathbf{y}$ is 
non-negative. Focusing on the first $4$ entries of $\mathbf{y}$, we have that
\begin{align}
	t_1 \ge 0, \\
	t_2q \ge t_1(n-(k+1)), \label{eqn:y_even_42}\\
	t_1\qty(\binom{n-(k+1)}{2} + q^{2}) + \frac{a_{k+3}}{q^{k+1}} \ge
	t_2q(n-(k+2)), \label{eqn:y_even_43}\\
	t_2q\binom{n-(k+2)}{2} + \frac{a_{k+4}}{q^{k+1}} \ge
	t_1\qty(\binom{n-(k+1)}{3}+q^2(n-(k+3))). \label{eqn:y_even_44}
\end{align}
Combining \autoref{eqn:y_even_42} and \autoref{eqn:y_even_43}, we have that
$$
t_1\qty(\binom{n-(k+1)}{2}+q^2) + \frac{a_{k+3}}{q^{k+1}}
\ge t_2q(n-(k+2)) \ge t_1(n-(k+1))(n-(k+2)),
$$
which implies
\begin{equation}
	\label{eqn:comb_even_1}
	\frac{a_{k+3}}{q^{k+1}} \ge t_1\qty(\binom{n-(k+1)}{2}-q^2).
\end{equation}
Multiplying $\frac{2}{n-(k+3)}$ on both sides of \autoref{eqn:y_even_44}, we have that
\begin{equation}
	\label{eqn:y_even_tmp}
	t_2q(n-(k+2)) + \frac{2a_{k+4}}{q^{k+1}(n-(k+3))} \ge
	t_1\qty(\frac{2}{3}\binom{n-(k+1)}{2}+2q^2).
\end{equation}
Combining \autoref{eqn:y_even_43} and \autoref{eqn:y_even_tmp}, we have
that
$$
t_1\qty(\binom{n-(k+1)}{2}+q^2) + \frac{a_{k+3}}{q^{k+1}}
\ge t_2q(n-(k+2)) \ge
t_1\qty(\frac{2}{3}\binom{n-(k+1)}{2}+2q^2) -
\frac{2a_{k+4}}{q^{k+1}(n-(k+3))},
$$
which implies
\begin{equation}
	\label{eqn:comb_even_2}
	t_1\qty(\frac{1}{3}\binom{n-(k+1)}{2}-q^2)
	\ge -\frac{a_{k+3}}{q^{k+1}} - \frac{2a_{k+4}}{(n-(k+3))q^{k+1}}.
\end{equation}
Combining \autoref{eqn:comb_even_1} and \autoref{eqn:comb_even_2},
we have that
\begin{equation*}
	\frac{a_{k+3}}{q^{k+1}} \ge t_1\qty(\binomial{n-(k+1)}{2}-q^2)
	\ge t_1\qty(\binomial{n-(k+1)}{2}-3q^2) \ge
	-\frac{3a_{k+3}}{q^{k+1}} - \frac{6a_{k+4}}{(n-(k+3))q^{k+1}},
\end{equation*}
which further implies
\begin{equation}
	\label{eqn:defect_2_even}
4a_{k+3} \ge - \frac{6a_{k+4}}{n-(k+3)}.
\end{equation}
Substituting $n = 2k+4$, $a_{k+3} $ and $a_{k+4} $ from \autoref{eqak34} 
into
\autoref{eqn:defect_2_even} and simplifying, we have that $3q^2 -1 \ge k$, which is a contradiction.
\end{proof}


\vspace{0.2cm}

\textbf{When $n$ is odd},  $n = 2k+5$.
In this case, $A_{k+1}^\prime = A_{k+4}^\prime$ and $A_{k+2}^\prime = A_{k+3}^\prime$
are unknown to us.
Similar to \autoref{eqn:linear_equation}, we rewrite  \autoref{eqn:k1k2k3k4_matrix} interms of all unknowns as follows
\begin{equation}
	\label{eqn:linear_equation_odd}
	\begin{pNiceArray}{cccc|cccc}
		\Block{4-4}<\Large>{M} &&&&-q^{k+1} & &\\
		&&&&& -q^{k+2}&\\
		&&&&& &-q^{k+3}\\
		&&&&  & &&-q^{k+4}\\
		\hline
		0 & 0 & 0 & 0          &1   &0    &0&-1\\
		0 & 0 & 0 & 0          &0   &1    &-1&0\\
	\end{pNiceArray}
	\spalignmat{
		A_{k+1};A_{k+2};A_{k+3};A_{k+4};A_{k+1}^\prime;A_{k+2}^\prime;A_{k+3}^\prime;A_{k+4}^\prime
	} =
	\spalignmat{
		-\binom{n}{k+1};-\binom{n}{k+2};-\binom{n}{k+3};-\binom{n}{k+4};0;0
	}.
\end{equation}
We still let $L$ denote the coefficient matrix in \autoref{eqn:linear_equation_odd}. Then
$\rank L = 6$. 
Define 
\begin{equation}
	\mathbf{y}_1 = \spalignmat{q^{k+1}\mathbf{v}_1;1;0;0;0}, \quad
	\mathbf{y}_2 = \spalignmat{q^{k+2}\mathbf{v}_2;0;1;0;0}, \quad
	\mathbf{y}_3 = \spalignmat{q^{k+3}\mathbf{v}_3;0;0;1;0}, \quad
	\mathbf{y}_4 = \spalignmat{q^{k+4}\mathbf{v}_4;0;0;0;1}.
\end{equation}
This time, $\mathbf{y}_1+\mathbf{y}_4$ and $\mathbf{y}_2+\mathbf{y}_3$ constitute a basis of the null space of $L$.
A particular solution is again cooked up from the enumerators of AME states.
Substituting
\begin{equation}
	a_{k+1}^\prime= a_{k+4}^\prime = \frac{\binom{n}{k+1}}{q^{k+1}}, \quad
	a_{k+2}^\prime = a_{k+3}^\prime = \frac{\binom{n}{k+2}}{q^{k+2}}.
\end{equation}
 into
\autoref{eqn:k1k2k3k4_matrix}, we get $a_{k+1} = a_{k+2} = 0$ and
\begin{equation}
	\label{eqn:enumerator_AME}
	a_{k+3} = (q-1)\binomial{n}{k+3}, \quad
	a_{k+4} = (q^3-1)\binom{n}{k+4} - (n-k-3)(q-1)\binomial{n}{k+3}.
\end{equation}
Then
$\mathbf{y}_p = (a_{k+1},a_{k+2},a_{k+3},a_{k+4},
a_{k+1}^\prime,a_{k+2}^\prime,a_{k+3}^\prime,a_{k+4}^\prime)^\intercal$
is a particular solution to
\autoref{eqn:linear_equation_odd}. So any solution to
\autoref{eqn:linear_equation_odd} is of the form
\begin{equation}\label{eqn:y_odd}
	\mathbf{y} = t_1(\mathbf{y}_1+\mathbf{y}_4) + t_2(\mathbf{y}_2+\mathbf{y}_3) + \mathbf{y}_p, \quad t_1,t_2 \in \RR.
\end{equation}
\begin{claim}\label{cl:l2odd}If $k > 3q^2 + 3q -1$, there must be a negative entry in $\mathbf{y}$ in \autoref{eqn:y_odd} for any choices of $t_1,t_2 \in \RR.$
\end{claim}
By \autoref{cl:l2odd}, when $n=2k+5>6q^2+6q+3$, $k$-uniform states in
$(\CC^q)^{\otimes n}$ do not exist, that is, 
\autoref{theorem:defect1} (2) follows for odd $n$. The proof of
\autoref{cl:l2odd} is similar to that of \autoref{cl:l2even}, which
can be found in Appendix \ref{sec:pf_l2_claim}.




\subsection{A Conjecture on Bounds of AME States of Defect $l$}
%
Motivated by the results in \autoref{theorem:Scott} and \autoref{theorem:defect1}, it is natural to raise the following
conjecture for the non-existence of AME states of defect $l$.
\begin{conjecture}
	\label{conj:ame_defect}
	Let $l \ge 0$ be a fixed integer. There exists a polynomial
	$f_l(q) = (2l+2)q^2 + o(q^2)$ such that if $n > f_l(q)$, then AME states of
	defect $l$ in $(\CC^q)^{\otimes n}$ do not exist.
\end{conjecture}

\section{Bounds from the Dual Linear Program}
\label{section:dual}
The bounds on AME states of defect $1$ or defect $2$ in \autoref{sec:defect1}
are derived by directly inspecting the linear constraints of
Eqs. \ref{eqn:unitary_constraints_1}--\ref{eqn:unitary_constraints_3}.
In this section, we turn to its dual linear program, 
and show bounds on general $k$-uniform states in  
\autoref{theorem:dual_unitary_supp_2}. 
For fixed $l \ge 3$, \autoref{theorem:dual_unitary_supp_2} implies  the 
non-existence of AME states of 
defect $l$ when $n\gtrapprox cq^2$ for some constant $c$ depending on $l$ (see \autoref{corcq}), which supports the truth 
of \autoref{conj:ame_defect}.
The asymptotic behavior of \autoref{theorem:dual_unitary_supp_2} provides the first bound of $k$ when $q=4$ and confirms \autoref{conj:q_4} (2) for $q=5$ in a stronger form (see \autoref{theorem:asymptotic} and thereafter).

\subsection{The Dual Linear Program}
First, we rewrite the linear program of
Eqs. \ref{eqn:unitary_constraints_1}--\ref{eqn:unitary_constraints_3}
more explicitly as follows.
Let
\begin{equation}
	M =
	\spalignmat{
		\binomial{n-k-1}{0} ;
		\binomial{n-k-1}{1} , \binomial{n-k-2}{0} ;
		\binomial{n-k-1}{2} , \binomial{n-k-2}{1} , \binomial{n-k-3}{0} ;
		\vdots , \vdots , \vdots , \ddots ;
		\binomial{n-k-1}{n-k-1} , \binomial{n-k-2}{n-k-2} ,
		\binomial{n-k-3}{n-k-3} , \dots , \binomial{0}{0}
	}.
\end{equation}
Then all equations in \autoref{eqn:unitary_constraints_3} can be written as
\begin{equation}
	M\spalignmat{
		A_{k+1};A_{k+2};\vdots;A_{n}
	} =
	\spalignmat{
		q^{k+1}A_{k+1}^\prime;q^{k+2}A_{k+2}^\prime;\vdots;q^nA_n^\prime
	}-\spalignmat{\binom{n}{k+1};\binom{n}{k+2};\vdots;\binom{n}{n}}.
\end{equation}
Let 
$\mathbf{m}^\intercal_{k+1},\mathbf{m}^\intercal_{k+2},\dots,\mathbf{m}^\intercal_{n}$ be the rows of
$M$, and let
$\mathbf{c}^\intercal = (0,\dots,0)_{n-k}$ be the zero vector of length $n-k$.
Let $\mathbf{x} = (x_{k+1},x_{k+2},\dots,x_n)^\intercal$ be a list of variables
of
length $n-k$,
and think of $x_i$ as $A_i$ for $i \in [k+1,n]$.
The linear program of
Eqs. \ref{eqn:unitary_constraints_1}--\ref{eqn:unitary_constraints_3}
can be written as
\begin{align}
	\text{minimize} \quad & \mathbf{c}^\intercal \mathbf{x} \label{eqn:obj}\\
	\text{subject to} \quad & \mathbf{m}_i^\intercal \mathbf{x} \ge 0 \defeq
	b_i, \quad
	i \in
	[k+1,\floor{n/2}], \label{eqn:c1}\\
	& \mathbf{m}_i^\intercal \mathbf{x} \ge \qty(q^{2i-n}-1)\binom{n}{i} \defeq
	b_i,
	\quad  i \in
	[\floor{n/2}+1,n-k-1], \label{eqn:c2}\\
	& \mathbf{m}_i^\intercal \mathbf{x} = \qty(q^{2i-n}-1)\binom{n}{i} \defeq
	b_i, \quad
	i \in
	[n-k,n], \label{eqn:c3} \\
	& x_i \ge 0, \quad i \in [k+1,n]. \label{eqn:c4}
\end{align}
Denote this linear program by $U(n,k,q)$.
Note that there are no objective functions contained in
Eqs. \ref{eqn:unitary_constraints_1}--\ref{eqn:unitary_constraints_3},
but here in $U(n,k,q)$, we artificially add an objective function 
(\autoref{eqn:obj}), which is $0$ as long as 
$U(n,k,q)$ has a feasible solution. The non-feasibility of
$U(n,k,q)$ will imply the non-existence of $k$-uniform states in
$(\CC^q)^{\otimes n}$ as before.

In the following, we present the dual linear program of $U(n,k,q)$, denoted by
$V(n,k,q)$, as in \cite[Chapter 4]{Bertsimas1997LPbook}. Let $\mathbf{M}_{k+1},\mathbf{M}_{k+2},\dots,\mathbf{M}_{n}$ be the columns of
$M$, and 
let $\mathbf{b} = (b_{k+1},b_{k+2},\dots,b_n)^\intercal$ be defined in Eqs. 
\ref{eqn:c1}--\ref{eqn:c3}. 
Setting $\mathbf{p} = (p_{k+1},\dots,p_{n})^\intercal$ to be the variables of
the dual
program, we can write $V(n,k,q)$ as
\begin{align}
	\text{maximize} \quad & \mathbf{p}^\intercal \mathbf{b},
	\label{eqn:dual_obj}\\
	\text{subject to} \quad & p_i \ge 0, \quad i \in [k+1,n-(k+1)],
	\label{eqn:dual_c1}\\
	& p_i \text{ free}, \quad  i \in [n-k,n], \label{eqn:dual_c2}\\
	& \mathbf{p}^\intercal \mathbf{M}_i  \le 0, \quad i \in [k+1,n].
	\label{eqn:dual_c3}
\end{align}
By \cite[Theorem 4.3]{Bertsimas1997LPbook}, if $\mathbf{x}$ is a feasible
solution to
$U(n,k,q)$, and $\mathbf{p}$ is a feasible solution to $V(n,k,q)$, then
$\mathbf{p}^\intercal \mathbf{b} \le \mathbf{c}^\intercal \mathbf{x} = 0$.
Equivalently, if we can find a solution $\mathbf{p}$ to $V(n,k,q)$ such that
$\mathbf{p}^\intercal \mathbf{b} > 0$, then $U(n,k,q)$ is not feasible, and thus
$k$-uniform states in $(\CC^q)^{\otimes n}$ do not exist.
We summarize this observation in the following lemma.
\begin{lemma}
	\label{lemma:dual_program}
Let $q \ge 2$, $k \le \floor{n/2}-1$, and
$p_{k+1},p_{k+2},\dots,p_n$ be real numbers such that
$p_{i} \ge 0$ for $i \in [k+1,n-(k+1)]$. If
\begin{equation}\label{eq:pm}
  \sum_{i=j}^n p_i \binom{n-j}{i-j} \le 0, \quad \forall j \in [k+1,n],
\end{equation}
and
\begin{equation}\label{eq:pb}
  \sum_{i = \floor{n/2}+1}^n p_i \qty(q^{2i-n}-1)\binom{n}{i} > 0,
\end{equation}
then $k$-uniform states in $(\CC^q)^{\otimes n}$ do not exist.
\end{lemma}
In the following, we will focus on constructing a solution
$\mathbf{p} = (p_{k+1},\dots,p_n)$
satisfying the conditions in \autoref{lemma:dual_program}.

\subsection{Bounds from Dual Solutions}
Let $u = \floor{n/2}+1$ and $v = n-k$. Assume that $u < v$. For simplicity, we
set $\mathbf{p} = (p_{k+1},\dots,p_{n})^\intercal$ with $p_u > 0$, $p_v = -1$
and all other entries zero. This setting satisfy
the easy constraint of $p$ in  \autoref{lemma:dual_program}. We next deduce the
sufficient conditions on the existence of $p_u$ so that $p$
satisfies all requirements in \autoref{lemma:dual_program}.
The constraints in \autoref{eq:pm} requires that
\begin{equation}
	\binom{n-k-1}{n-u}p_u \le \binom{n-k-1}{n-v},
	\binom{n-k-2}{n-u}p_u \le \binom{n-k-2}{n-v},
	\dots,\binom{n-u}{n-u}p_u \le \binom{n-u}{n-v}.
\end{equation}
As $\frac{\binom{x}{n-v}}{\binom{x}{n-u}}$ is decreasing with respect to $x$,
it suffices that
\begin{equation}
	\label{eqn:dual_solution_1}
	p_u \le \frac{\binom{n-k-1}{n-v}}{\binom{n-k-1}{n-u}}.
\end{equation}
On the other hand, \autoref{eq:pb}  is equivalent to
\begin{equation}
	\label{eqn:dual_solution_2}
	p_{u} >
	\frac{(q^{2v-n}-1)\binom{n}{v}}{(q^{2u-n}-1)\binom{n}{u}}.
\end{equation}
%
So if the inequalities in \autoref{eqn:dual_solution_1} and \autoref{eqn:dual_solution_2} have a common solution $p_u$,
then there exists a solution $p$ to $V(n,k,q)$
satisfying the conditions in \autoref{lemma:dual_program}.
We conclude in the following theorem.
\begin{theorem}
	\label{theorem:dual_unitary_supp_2}
	Let $k \le \floor{n/2}-1$, $u = \floor{n/2}+1$ and $v = n-k$.
	If $u<v$ and
\begin{equation}\label{eqthm}
	\frac{\binom{n-k-1}{n-v}}{\binom{n-k-1}{n-u}}>
	\frac{(q^{2v-n}-1)\binom{n}{v}}{(q^{2u-n}-1)\binom{n}{u}},
\end{equation}
then $k$-uniform states in
	$(\CC^q)^{\otimes n}$ do not exist.
\end{theorem}

We have the following immediate consequence by considering even $n$ and odd $n$ in \autoref{theorem:dual_unitary_supp_2} explicitly.

\begin{corollary}\label{theorem:numerical}
Suppose that $l\geq 1$ when $n$ is odd, and $l\geq 2$ when $n$ is even.
 Then AME states of defect
$l$ in $(\CC^q)^{\otimes n}$ do not exist if
\begin{equation}\label{eqn:numerical_0}
\begin{cases}
	\frac{\factorial{(m+l)}}{\factorial{(m+1)}} >
	\frac{q^{2l}-1}{q^2-1} \cdot
	\frac{\factorial{(2l-1)}}{\factorial{l}}, & n = 2m,
	\\
	\frac{\factorial{(m+l+1)}}{\factorial{(m+1)}} >
	\frac{q^{2l+1}-1}{q-1} \cdot
	\frac{\factorial{(2l)}}{\factorial{l}}, & n = 2m+1.
\end{cases}
\end{equation}
\end{corollary}

For fixed $l$, the left-hand side of the inequalities in
\autoref{eqn:numerical_0} is increasing with respect to $m$. So 
\autoref{theorem:numerical} provides a Scott-type bound of AME states of defect
$l$ for general $l$. Now we compare the bounds for $l=1,2$ between \autoref{theorem:numerical} and \autoref{theorem:defect1}. When $l=1$ and $n$ is odd, or when $l=2$ and $n$ is even, the bounds are the same.
When $l=2$ and $n$ is odd, the bound by \autoref{theorem:numerical} is stronger when $q \le 3$, but weaker when $q \ge 5$, and  the same when $q=4$.
%
For $l = 3,4$ and $q = 4,5,6,7$,  the corresponding
non-existence results of AME states of defect $l$ by 
\autoref{theorem:numerical} are listed in \autoref{tab:ame_small_defects}.

\setcounter{table}{2}
\begin{table}
	\centering
	\caption{Non-Existence of AME States of Defect $l \in \setnd{3,4}$
		in $(\CC^q)^{\otimes n}$}
	\label{tab:ame_small_defects}
	\begin{threeparttable}
		\begin{subtable}{.3\textwidth}
			\begin{tabular}{|c|c|c|}
				\toprule
				\multicolumn{3}{|c|}{$q = 4$} \\
				\midrule
				$l$ & $n$ even & $n$ odd  \\
				\midrule
				$3$ & $n \ge 144$ & $n \ge 169$ \\
				$4$ & $n \ge 190$ & $n \ge 215$ \\
				\bottomrule
			\end{tabular}
		\end{subtable}
		\begin{subtable}{.3\textwidth}
			\begin{tabular}{|c|c|c|}
				\toprule
				\multicolumn{3}{|c|}{$q = 5$} \\
				\midrule
				$l$ & $n$ even & $n$ odd  \\
				\midrule
				$3$ & $n \ge 224$ & $n \ge 261$ \\
				$4$ & $n \ge 296$ & $n \ge 333$ \\
				\bottomrule
			\end{tabular}
		\end{subtable}
		\begin{subtable}{.3\textwidth}
			\begin{tabular}{|c|c|c|}
				\toprule
				\multicolumn{3}{|c|}{$q = 6$} \\
				\midrule
				$l$ & $n$ even & $n$ odd  \\
				\midrule
				$3$ & $n \ge 322$ & $n \ge 373$ \\
				$4$ & $n \ge 428$ & $n \ge 477$ \\
				\bottomrule
			\end{tabular}
		\end{subtable}
		\begin{subtable}{.3\textwidth}
			\begin{tabular}{|c|c|c|}
				\toprule
				\multicolumn{3}{|c|}{$q = 7$} \\
				\midrule
				$l$ & $n$ even & $n$ odd  \\
				\midrule
				$3$ & $n \ge 438$ & $n \ge 505$ \\
				$4$ & $n \ge 582$ & $n \ge 647$ \\
				\bottomrule
			\end{tabular}
		\end{subtable}
	\end{threeparttable}
\end{table}

\autoref{theorem:numerical} is also a step forward to \autoref{conj:ame_defect}, which can be seen from the following consequence.
\begin{corollary}\label{corcq}
  For fixed $l \ge 2$, there is a constant $c$ depending only on
$l$ such that for $q$ large enough and $n > cq^2$, AME states of defect $l$ in $(\CC^q)^{\otimes n}$ do not exist.
\end{corollary}
This confirms \autoref{conj:ame_defect} for the order of $q$, but leaves the 
leading coefficient undetermined.

By an asymptotic analysis of \autoref{theorem:dual_unitary_supp_2},
we have the following asymptotic bound on $k$-uniform states,
whose proof can be found in Appendix \ref{sec:asymptotic}.
\begin{theorem}
	\label{theorem:asymptotic}
	Let $\theta \le 1/2$ be such that
	\begin{equation}\label{eq:theta}
	  	\frac{1}{2}\log(1-2\theta) +
	(1-\theta)\log\qty(\frac{1-\theta}{1-2\theta})+
	\theta\log(2) - (1-2\theta) \log q > 0.
	\end{equation}
	Then, for $k$-uniform states in $(\CC^q)^{\otimes n}$, we have that
	$$
	k \le \theta n
	$$
	for $n$ large enough.
\end{theorem}
We list several upper bounds of $k/n$ in \autoref{tab:asymptotic} for $k$-uniform states in $(\CC^q)^{\otimes n}$ by finding some suitable $\theta$ satisfying \autoref{eq:theta}. For $q=2,3$, we could not get a better bound than \autoref{theorem:Rains} from  \autoref{theorem:asymptotic}.
For $q = 4$, we get the first non-trivial upper bound of $k/n\leq 0.479$, but
still has a gap to \autoref{conj:q_4} (1).  For $q=5$, the upper bound is
$k/n\leq 0.487$ in \autoref{tab:asymptotic}, which confirms
\autoref{conj:q_4} (2) asymptotically in a stronger form.


\setcounter{table}{3}
\begin{table}
	\centering
	\caption{Asymptotic Bounds on $k$-Uniform States in $(\CC^q)^{\otimes n}$}
	\label{tab:asymptotic}
	\begin{tabular}{c|cccccc}
		\toprule
		$q$       & $4$ & $5$ & $6$ & $7$ & $8$ & $9$   \\
		\midrule
		$\frac{k}{n} \le$ & $0.479$ & $0.487$ & $0.491$ & $0.494$ & $0.495$ &
		$0.496$ \\
		\bottomrule
	\end{tabular}
\end{table}

\section{Improvement on the Shadow Bounds}
\label{sec:extremal}
In this section, we aim to improve the shadow bounds in \autoref{theorem:Rains} and \autoref{conj:q_4}  by a constant
for $q = 3,4,5$. For $q = 2$, the improvement on Rains' shadow bound
(\autoref{theorem:Rains} (1)) by a constant has been considered in
\cite{Han2006Nonexistence,Han2009Nonexistence}.

Our method is a refinement of  the
method to the shadow bounds. We first propose a criterion
on the non-existence of $k$-uniform states (see \autoref{prop:extremal}).
Based on this criterion, we provide with numerical results which improve
the shadow bounds by a constant (see \autoref{tab:extremal_3},
\autoref{tab:extremal_4} and \autoref{tab:extremal_5}).
For $q=4$, our results  improve some of the bounds in the code tables
\cite{Grassl:codetables} (see \autoref{tab:improvement_4}). Strictly speaking, the bounds in the code tables \cite{Grassl:codetables}
only applies to additive codes, while our bounds applies to any codes.

\subsection{Review of the Approach to the Shadow Bounds}
\label{subsec:review}
We review the proof of the shadow bounds, by setting up
some notations and introducing some necessary results.
For more details, we refer to \cite{Rains1998Shadow,Shi2023} and
the references therein.

Let $\ket{\psi}$ be a pure state in $(\CC^q)^{\otimes n}$. Let $A(x,y) =
\sum_{j=0}^n A_jx^{n-j}y^j$ be the weight enumerator  associated with
$\ket{\psi}$.
By a Gleason-type theorem \cite{MacWilliams1972Gleason},
$A(x,y)$ is a polynomial in $x+(q-1)y$ and $y(x-y)$,
so we may write (see \cite[Eq. (16)]{Shi2023})
\begin{equation}
	\label{eqn:Gleason_A}
	A(x,y) = \sum_{i=0}^{\floor{n/2}}c_i(x+(q-1)y)^{n-2i}(y(x-y))^i,
\end{equation}
where $c_i \in \RR, i \in [0,\floor{n/2}]$. 
Relating \autoref{eqn:Gleason_A} with the weight enumerator, we can write
\begin{equation}
	\label{eqn:A_to_c}
	c_i = \sum_{j=0}^i \alpha_{i,j}A_j, \quad 0 \le i \le \floor{n/2},
\end{equation}
where $\alpha_{i,j}$ are some real numbers (see \cite[Eq. (20)]{Shi2023}).
We know that
\begin{equation}
	\label{eqn:alpha_diagonal}
\alpha_{i,i} = 1, \quad i \in [0,\floor{n/2}].
\end{equation}
For $i \in [1,\floor{n/2}]$, $\alpha_{i,0}$ can be expressed as
(see \cite[Appendix B and Lemma 5]{Shi2023})
\begin{equation}\label{eqn:alphai02}
\begin{aligned}
	\alpha_{i,0} &= -\frac{n(q-1)}{i}\qty[\text{coeff. of } y^{i-1} \text{ in }
	(1+(q-1)y)^{-(n+1-2i)}(1-y)^{-i}],\\
	&= -\frac{n(q-1)}{i} \sum_{j=0}^{i-1}(1-q)^j \binomial{n-2i+j}{n-2i}
	\binomial{2i-2-j}{i-1}.
\end{aligned}
\end{equation}

Let $S(x,y) = \sum_{j=0}^n S_jx^{n-j}y^j$ be the shadow enumerator associated with $\ket{\psi}$. As $S(x,y) = A\qty(\frac{(q-1)x+(q+1)y}{q},\frac{y-x}{q})$, we have that
\begin{equation}
	\label{eqn:Gleason_S}
	S(x,y) = \sum_{i=0}^{\floor{n/2}}(-1)^ic_i2^{n-2i}q^{-i}y^{n-2i}(x^2-y^2)^i.
\end{equation}
By comparing coefficients between \autoref{eqn:Gleason_S} and
the shadow enumerator,
 \cite[Lemma 6]{Shi2023} shows that
\begin{equation}
	\label{eqn:sign}
	(-1)^ic_i \ge 0 \text{ for all } 0 \le i \le \floor{n/2}.
\end{equation}

The shadow bounds in \autoref{theorem:Rains}
can be derived by determining the smallest $i \le \floor{n/2}$ such that  $(-1)^ic_{i}<0$.
In fact, if $\ket{\psi}$ is a $k$-uniform state for $k\geq i$, then
$A_0 = 1$ and $A_1=A_2=\dots=A_{i} = 0$. Thus, by \autoref{eqn:A_to_c}, $c_i = \alpha_{i,0}$.
If $(-1)^i \alpha_{i,0} < 0$, we have a
contradiction to \autoref{eqn:sign}, which implies the non-existence of $k$-uniform states
(see \cite[Lemma 7]{Shi2023}), and thus $k\leq i-1$. The difficulty
lies in determining the sign of $\alpha_{i,0}$ for  $i \in [1,\floor{n/2}]$ in 
\autoref{eqn:alphai02}. The shadow bounds in \autoref{conj:q_4} were 
verified for small $n$ by checking the sign of $\alpha_{i,0}$ numerically in 
\cite{Shi2023}.


\subsection{Refinement of the Shadow Bounds}
The non-existence of $k$-uniform states in \autoref{theorem:Rains}
are derived by inspecting the sign of  $c_{k}$ as discussed before.
In this subsection, we further inspect $c_{k+1}$ and $c_{k+2}$ to
strengthen the shadow bounds.

Suppose that $k \le \floor{n/2}-2$ is odd and $\ket{\psi}$ is a $k$-uniform
state in $(\CC^q)^{\otimes n}$. Then $A_0=1$ and
$A_1=A_2=\dots=A_k=0$.
By \autoref{eqn:A_to_c} and \autoref{eqn:alpha_diagonal}, we have
\begin{align}
	c_{k+1} &= \alpha_{k+1,0} + A_{k+1}, \label{eqn:c_k1}\\
	c_{k+2} &= \alpha_{k+2,0} + \alpha_{k+2,k+1}A_{k+1} + A_{k+2}
	\label{eqn:c_k2}.
\end{align}
Since $k$ is odd, by \autoref{eqn:sign}, $c_{k+1}\geq 0$ and $c_{k+2}\leq 0$.
Combining \autoref{eqn:c_k1} and
\autoref{eqn:c_k2}, we have that
\begin{align}
	A_{k+2} &\le -\alpha_{k+2,0} - \alpha_{k+2,k+1}A_{k+1} \label{eqn:ub1} \\
	&\le -\alpha_{k+2,0} + \alpha_{k+2,k+1}\alpha_{k+1,0}, \label{eqn:ub2}
\end{align}
where the inequality in \autoref{eqn:ub2} is based on the assumption that
$\alpha_{k+2,k+1} \ge 0$. As $A_{k+2} \ge 0$, if furthermore
$-\alpha_{k+2,0} + \alpha_{k+2,k+1}\alpha_{k+1,0} < 0$, we will have a
contradiction, implying the non-existence of the pure $k$-uniform state
$\ket{\psi}$ in $(\CC^q)^{\otimes n}$. The value of $\alpha_{k+2,k+1}$ can be computed in the following lemma, whose proof can be
found
in Appendix \ref{sec:proof_alpha_off_diagonal} .
\begin{lemma}
	\label{lemma:alpha_off_diagonal}
	$\alpha_{k+2,k+1} = (k+1)(2q-1)-(q- 1)n$.
\end{lemma}

We summarize the above argument in the following proposition.
\begin{proposition}
	\label{prop:extremal}
	Let $k$ be odd and $k \le \floor{n/2}-2$. Let $\alpha_{i,0}$  be defined in
 \autoref{eqn:alphai02} for $i \in [1,\floor{n/2}]$. If $(k+1)(2q-1)-(q- 1)n \ge 0$ and
	\begin{equation}
		\label{eqn:extremal_condition}
	 ((k+1)(2q-1)-(q- 1)n)\alpha_{k+1,0} < \alpha_{k+2,0},
	\end{equation}
	then $k$-uniform states in $(\CC^q)^{\otimes n}$ do not exist.
\end{proposition}

\subsection{Numerical Results}
In this subsection, we provide some numerical results
by checking  the conditions in \autoref{prop:extremal}. For convenience, let $S_q(n)$ denote the
value of the shadow bounds for $k$-uniform states in $(\CC^q)^{\otimes n}$ listed in \autoref{theorem:Rains} and \autoref{conj:q_4}. For $q=2$, the improvement on Rains' shadow bound
(\autoref{theorem:Rains} (1)) by a constant has been studied in
\cite{Han2006Nonexistence,Han2009Nonexistence}.

\subsubsection{$q=3$}
Write $n = 14m + r$  with $r \in [0,13]$.
For $m \in [1,200]$, we checked the conditions of \autoref{prop:extremal} 
numerically to show the non-existence of $(S_3(n)-l)$-uniform states in 
$(\CC^3)^{\otimes n}$, where $l \in \setnd{0,2,4,6,8}$. This gives the upper 
bound $k \le S_3(n) - l -1$ for $k$-uniform states in $(\CC^3)^{\otimes n}$.
Detailed results are listed in \autoref{tab:extremal_3}.

\begin{table}
	\centering
	\caption{Improvement on Shadow Bounds for
		$k$-Uniform States in $(\CC^3)^{\otimes n}$, $n=14m+r$}
	\label{tab:extremal_3}
\begin{threeparttable}
\begin{tabular}{c|ccccccc}
	\toprule
	$m \ge $& $r=0$ & $r=1$ &$r=2$ &$r=3$ &$r=4$ &$r=5$ &$r=6$ \\
	\midrule
	$l=0$ & $3$ & $4$ & $6$ & $7$ & $10$ & $3$ & $5$ \\
	\midrule
	$l=2$ & $14$ & $23$ & $34$ & $45$ & $57$ & $16$ & $26$ \\
	\midrule
	$l=4$ & $65$ & $76$ & $88$ & $100$ & $111$ & $68$ & $80$  \\
	\midrule
	$l=6$ & $119$ & $131$ & $143$ & $155$ & $166$ & $123$ & $135$ \\
	\midrule
	$l=8$ & $175$ & $186$ & $198$ & --&-- & $178$ & $190$ \\
	\bottomrule
\end{tabular}

\bigskip
\begin{tabular}{c|ccccccc}
	\toprule
	$m \ge $& $r=7$ &$r=8$ 	&$r=9$ 	&$r=10$ &$r=11$ &$r=12$ &$r=13$ \\
	\midrule
	$l=0$ & $6$ & $7$ & $11$ & $3$ & $5$ & $6$ & $8$ \\
	\midrule
	$l=2$ & $37$ & $48$ & $60$ & $19$ & $29$ & $41$ & $52$ \\
	\midrule
	$l=4$ & $91$ & $103$ & $115$ & $72$ & $83$ & $95$ & $107$ \\
	\midrule
	$l=6$ & $147$ & $158$ & $170$ & $127$ & $138$ & $150$ & $162$ \\
	\midrule
	$l=8$ & --& --& --& $182$ & $194$ & --& --\\
	\bottomrule
\end{tabular}
\begin{tablenotes}
\footnotesize
\item This table can be read as follows. For $r$ and $l$, we have the
non-existence of $(S_3(n)-l)$-uniform
states in $(\CC^3)^{\otimes n}$, where  $n = 14m + r$ with $m$ between the
$(r,l)$-entry and $200$. This gives an upper bound $k \le S_3(n)-l-1$ for such
$n$. For example, when $r=0$ and $l=0$, we have $k \le S_3(n)-1=6m$ for $n=14m$
with $m\in[3,200]$, reducing the corresponding shadow bound by one. 
\end{tablenotes}
\end{threeparttable}
\end{table}

For $n \in [1,100]$, the online code tables \cite{Grassl:codetables} lists
upper bounds on the minimum distance $d$ for $[[n,0,d]]_3$ additive codes, which
correspond to $(d-1)$-uniform stabilizer states in $(\CC^3)^{\otimes n}$.
We note that our results listed in \autoref{tab:extremal_3}, coincide with
the results listed in the code tables, whenever $n=14m+r \le 97$. Note that the
code tables only provided with results for $n \in [1,100]$ when $q = 3$.
We provide  much more results for $n \in[101,2813]$.

\subsubsection{$q=4$} In this case, the values of $S_4(n)$ is from \autoref{conj:q_4} (1), and the correctness of this bounds has been confirmed in \cite{Shi2023} for $n\in [80,161]$.

Write $n = 17m + r$  with $r \in [0,16]$.
For $m \in [1,200]$, we checked the conditions of \autoref{prop:extremal} 
numerically to show the non-existence of $(S_4(n)-l)$-uniform states in 
$(\CC^4)^{\otimes n}$, where $l \in \setnd{0,2,4,6,8,10}$. 
Detailed results are listed in \autoref{tab:extremal_4}. Here $n$ is in the 
interval $[66,3416]$.



\begin{table}
	\centering
	\caption{Improvement on Shadow Bounds for
		$k$-Uniform States in $(\CC^4)^{\otimes n}$, $n = 17m+r$}
	\label{tab:extremal_4}
	\begin{threeparttable}
\begin{tabular}{c|ccccccccc}
\toprule
$m \ge $& $r=0$ & $r=1$ &$r=2$ &$r=3$ &$r=4$ &$r=5$ &$r=6$ &$r=7$ &$r=8$ \\
\midrule
$l=0$ & $6$ & $5$ & $4$ & $5$ & $7$ & $5$ & $4$ & $4$ & $6$ \\
\midrule
$l=2$ & $9$ & $17$ & $27$ & $36$ & $46$ & $15$ & $24$ & $34$ & $43$
\\
\midrule
$l=4$ & $49$ & $59$ & $68$ & $78$ & $88$ & $56$ & $66$ & $76$ & $85$  \\
\midrule
  $l=6$ & $91$ & $101$ & $111$ & $121$ & $130$ & $98$ & $108$ & $118$ & $128$ \\
\midrule
  $l=8$ & $133$ & $143$ & $153$ & $163$  & $173$ & $140$ & $150$  & $160$ &
  $170$\\
\midrule 
$l=10$ & $175$ & $185$ & $195$ & -- & -- & $183$ & $192$ & -- & -- \\
\bottomrule
\end{tabular}
		
		\bigskip
\begin{tabular}{c|cccccccc}
\toprule
$m \ge $& $r=9$ &$r=10$ & $r=11$ & $r=12$ &$r=13$ &$r=14$ &$r=15$ & $r=16$ \\
\midrule
  $l=0$ & $5$ & $4$ & $4$ & $5$ & $5$ & $4$ & $3$ & $4$ \\
\midrule
  $l=2$ & $12$ & $21$ & $31$ & $41$ & $10$ & $19$ & $28$ & $38$\\
\midrule
  $l=4$ & $53$ & $63$ & $73$ & $83$ & $51$ & $60$ & $70$ & $80$\\
\midrule
  $l=6$ & $95$ & $105$ & $115$ & $125$ & $93$ & $103$ & $112$ & $122$\\
\midrule
  $l=8$ & $138$ & $147$ & $157$ & $167$ & $135$ & $145$ & $155$ & $165$\\
\midrule 
  $l=10$ & $180$ & $190$ & $200$ & -- & $177$ & $187$ & $197$ & -- \\
\bottomrule
\end{tabular}
\begin{tablenotes}
\footnotesize
\item This table can be read in a fashion simliar to that of
\autoref{tab:extremal_3}. 
\end{tablenotes}
\end{threeparttable}
\end{table}

We note that some of the results listed in \autoref{tab:extremal_4} also
improve the results in the code tables \cite{Grassl:codetables}, which we
compare in \autoref{tab:improvement_4}.

\begin{table}

\end{table}

\begin{table}
\centering
\caption{New Upper Bounds on the Minimum Distance of $((n,1,d))_4$ Quantum
Codes}
\label{tab:improvement_4}
\begin{threeparttable}
\begin{tabular}{lcccccc}
	\toprule
	$n$ & $82,83$ & $86,87,88$ & $91,92$ & $96,97$ & $100$& \text{references} \\
	\midrule
	$d \le $& $39$ & $41$ & $43$
	& $45$ & $47$ &  \autoref{tab:extremal_4}\\
	\midrule
	$d \le $ & $40$
	& $42$ & $44$ & $46$ & $48$ &  code tables in \cite{Grassl:codetables}\\
	\bottomrule
\end{tabular}
\begin{tablenotes}
\footnotesize
\item Note that we are listing the upper bounds of the minimum distance
$d$ of $((n,1,d))_4$ quantum codes, which are equivalent to
$(d-1)$-uniform states in $(\CC^4)^{\otimes n}$. 

\end{tablenotes}
\end{threeparttable}
\end{table}

\subsubsection{$q=5$} We write $n=4m+r$ with $r\in [0,3]$. In this case, the values of $S_5(n)$ is from \autoref{conj:q_4} (2), and the correctness of this bounds has been confirmed in \cite{Shi2023} for $n\in [180,276]$.
%
%

\autoref{tab:extremal_5} listed some new bounds on $k$-uniform states in
$(\CC^5)^{\otimes n}$ for $n\in [180, 803]$, which were confirmed numerically 
with \autoref{prop:extremal}.

\begin{table}
	\centering
	\caption{Improvement on Shadow Bounds for
		$k$-Uniform States in $(\CC^5)^{\otimes n}$, $n = 4m+r$}
	\label{tab:extremal_5}
	\begin{threeparttable}
	\begin{tabular}{c|cccc}
		\toprule
		$m \ge $& $r=0$ &$r=1$ & $r=2$ & $r=3$ \\
		\midrule
		$l=0$ & -- & -- & $45$ & $46$ \\
		\midrule
		$l=2$ & $59$ & $71$ & $83$ & $95$ \\
		\midrule
		$l=4$ & $108$ & $121$ & $133$ & $145$\\
		\midrule
		$l=6$ & $158$ & $171$ & $183$ & $195$ \\
		\bottomrule
	\end{tabular}
\begin{tablenotes}
\footnotesize
\item This table can be read in a similar fashion to that of
\autoref{tab:extremal_3} and \autoref{tab:extremal_4}.
\end{tablenotes}
\end{threeparttable}
\end{table}

Note that although we only confirmed the non-existence results for finitely many $n$ when $q=3,4,5$, 
we strongly believe that the same bound holds for all larger $n$ in the 
corresponding congruent class. 
However, the proof requires some clever analysis, which we haven't found yet.

\section{Conclusion}
\label{sec:conclusion}
In this paper, we have studied the linear programming bounds on $k$-uniform
states, and provided several improvements on the bounds of $k$.
There are still some questions that might be interesting. For example,
does \autoref{conj:ame_defect} really hold?
Also, the dual solution in \autoref{section:dual}-B is a simple one, but
not necessarily a good one. Can better dual solutions be proposed to
improve the bounds in \autoref{section:dual}, or to show \autoref{conj:q_4}
asymptotically? Finally, can  \autoref{prop:extremal} be applied for infinitely many $n$?

\appendices
\section{A Proof of \autoref{cl:l2odd}}
\label{sec:pf_l2_claim}
The proof is again by contradiction and we assume that there is a choice of 
$t_1,t_2 \in \RR$ such that each entry of $\mathbf{y}$ in \autoref{eqn:y_odd}
is non-negative. Focusing on the first $4$ entries of $\mathbf{y}$, we have that
\begin{align}
	t_1 &\ge 0, \label{eqn:y_odd_41}\\
	t_2q &\ge t_1(n-k-1), \label{eqn:y_odd_42}\\
	t_1\binom{n-k-1}{2} + \frac{a_{k+3}}{q^{k+1}} &\ge t_2q(n-k-2-q),
	\label{eqn:y_odd_43}\\
	t_2q\pqty{\binom{n-k-2}{2}-q(n-k-3)}+\frac{a_{k+4}}{q^{k+1}} &\ge
	t_1\pqty{\binom{n-k-1}{3}-q^3}. \label{eqn:y_odd_44}
\end{align}
Combining \autoref{eqn:y_odd_42} and \autoref{eqn:y_odd_43}, we have
that
\begin{align*}
	t_2q\frac{n-k-2}{2} \ge t_1\binomial{n-k-1}{2} \ge
	t_2q(n-k-2-q) - \frac{a_{k+3}}{q^{k+1}},
\end{align*}
which is simplified as
\begin{equation}
	\label{eqn:comb_odd_1}
	\frac{a_{k+3}}{q^{k+1}} \ge t_2q\qty(\frac{n-k-2}{2} - q).
\end{equation}
Multiplying $\frac{3}{n-k-3}$ on both sides of \autoref{eqn:y_odd_44},
we have that
\begin{equation}
	\label{eqn:y_odd_44_simplified}
	3t_2q\qty(\frac{n-k-2}{2}-q) + \frac{3a_{k+4}}{(n-k-3)q^{k+1}} \ge
	t_1 \qty(\binomial{n-k-1}{2}-\frac{3q^3}{n-k-3}).
\end{equation}
By \autoref{eqn:y_odd_42}, we have that
\begin{equation}
	\label{eqn:y_odd_42_1}
	-t_1 \frac{3q^3}{n-k-3} \ge -\frac{3t_2q^4}{(n-k-1)(n-k-3)}.
\end{equation}
Substituting \autoref{eqn:y_odd_42_1} and \autoref{eqn:y_odd_43} into
\autoref{eqn:y_odd_44_simplified}, we have that
$$
3t_2q\qty(\frac{n-k-2}{2} -q) + \frac{3a_{k+4}}{(n-k-3)q^{k+1}} \ge
t_2q(n-k-2-q) - \frac{a_{k+3}}{q^{k+1}} -
\frac{3t_2q^4}{(n-k-1)(n-k-3)},
$$
which is further simplified as
\begin{equation}
	\label{eqn:comb_odd_2}	
	t_2q\qty(\frac{n-k-2}{2} - 2q + \frac{3q^3}{(n-k-1)(n-k-3)}) \ge
	-\frac{a_{k+3}}{q^{k+1}} - \frac{3a_{k+4}}{(n-k-3)q^{k+1}}.
\end{equation}
Note that $t_2 \ge 0$ by \autoref{eqn:y_odd_41} and
\autoref{eqn:y_odd_42}.
By the assumption that $k > 3q^2+3q-1$ in \autoref{cl:l2odd}, we have that 
\begin{equation}
	\label{eqn:defect2_odd_precondition}
	\frac{3q^3}{(n-k-1)(n-k-3)} \le q.
\end{equation}
Then, combining \autoref{eqn:comb_odd_1} and \autoref{eqn:comb_odd_2},
we
have that
\begin{equation}
	\label{eqn:defect_2_odd}
	\frac{a_{k+3}}{q^{k+1}} \ge -\frac{a_{k+3}}{q^{k+1}} -
	\frac{3a_{k+4}}{(n-k-3)q^{k+1}}.
\end{equation}
Substituing the values of $n = 2k+5$ and $a_{k+3},a_{k+4}$ into
\autoref{eqn:defect_2_odd} and simplifying, we have that $k \le 3q^2 +
3q - 1$, which is a contradiction to the assumption of
\autoref{cl:l2odd}.

\section{A Proof of \autoref{theorem:asymptotic}}
\label{sec:asymptotic}
First, we recall Stirling's approximation \cite{Robbins1955Remark}:
\begin{equation}
	\label{eqn:Stirling}
	\frac{1}{12n+1} < \log (\factorial{n}) - \qty[
	\qty(n+\frac{1}{2})\log n - n + \frac{1}{2}\log(2\pi)
	] < \frac{1}{12n}, \quad n = 1,2,3,\dots
\end{equation}
where $\log$ denotes the natural logarithm.

Now, we follow the notations of
\autoref{theorem:dual_unitary_supp_2} and show \autoref{theorem:asymptotic}.
Simplifying the condition in \autoref{eqthm}, we have that
\begin{equation}
	\label{eqn:fnkq_simplified}
	\frac{\factorial{v}\factorial{(u-k-1)}}{\factorial{u}\factorial{(n-2k-1)}}
	=
	\frac{\binomial{n-k-1}{n-v}\binomial{n}{u}}{\binomial{n-k-1}{n-u}\binomial{n}{v}}
	 >
	\frac{q^{2v-n}-1}{q^{2u-n}-1}.
\end{equation}
Note that $2u-n$ equals $1$ or $2$, depending on the parity of $n$.
Taking the natural logarithm, for \autoref{eqn:fnkq_simplified} to hold, it
suffices that
\begin{equation}
	\label{eqn:fnkq_suffice}
	\log(\factorial{v}) + \log(\factorial{(u-k-1)}) - \log(\factorial{u}) -
	\log(\factorial{(n-2k-1)}) \ge (2v-n)\log q.
\end{equation}
By Stirling's approximation, \autoref{eqn:fnkq_suffice} holds if
\begin{multline}
	\label{eqn:fnkq_suffice_stirling}
	(v+1/2)\log(v) + (u-k-1/2)\log(u-k-1) -  \\
	(u+1/2)\log(u)
	-(n-2k-1/2)\log(n-2k-1) \ge (2v-n)\log q.
\end{multline}
Let $\theta = \frac{k}{n}$. \autoref{eqn:fnkq_suffice_stirling} holds
asymptotically, as long as
the following equation holds asymptotically:
\begin{multline}
	\label{eqn:fnkq}
	(1-\theta)\log((1-\theta)n) + (1/2-\theta)\log((1/2-\theta)n)  \\
	- \frac{1}{2}\log(n/2) - (1-2\theta)\log((1-2\theta)n) - \order{\frac{\log
	n}{n}} > (1-2\theta)\log q.
\end{multline}
The left-hand side of \autoref{eqn:fnkq} simplifies as
\begin{equation}
	\frac{1}{2}\log\qty(
	\frac{(1/2-\theta)n}{n/2}
	)
	+(1-\theta) \log\qty(
	\frac{(1-\theta)n}{(1-2\theta)n}
	)
	+\theta\log\qty(
	\frac{(1-2\theta)n}{(1/2-\theta)n}
	)-\order{\frac{\log n}{n}},
\end{equation}
which converges to
\begin{equation}
	\frac{1}{2}\log(1-2\theta) +
	(1-\theta)\log\qty(\frac{1-\theta}{1-2\theta})+
	\theta\log(2)
\end{equation} as $n$ goes to infinity.
In summary, for \autoref{theorem:dual_unitary_supp_2} to hold asymptotically, it
suffices that
\begin{equation}
	\frac{1}{2}\log(1-2\theta) +
	(1-\theta)\log\qty(\frac{1-\theta}{1-2\theta})+
	\theta\log(2) - (1-2\theta) \log q > 0.
\end{equation}
The proof is completed as desired.

\section{A Proof of \autoref{lemma:alpha_off_diagonal}}
\label{sec:proof_alpha_off_diagonal}
Let $f(y)$ be a formal power series, and we use the notation $f[i]$ to denote
the coefficient of $y^i$ in $f(y)$ for integers $i \ge 0$. The proof of
\autoref{lemma:alpha_off_diagonal} is a routine application of the
Lagrange-B\"{u}rmann theorem.
\begin{lemma}[Lagrange-B\"{u}rmann
\cite{Shi2023,Rains1998Shadow,Mallows1973bound}]
	\label{lemma:LagrangeBurmann}
	Let $f(y)$ and $g(y)$ be formal power series. Assume that $g(0) = 0$ and
	$g^\prime(0) \neq 0$.
	Then $f(y)$ can be expanded as a series in $g(y)$ as
	$$
	f(y) = \sum_{i \ge 0}k_i g(y)^i,
	$$
	where
	$$
	k_i = \frac{1}{i}\qty[f^\prime(y)\qty(\frac{y}{g(y)})^i][i-1],
	\quad i \ge 1.
	$$
\end{lemma}
Now, we proceed with the proof of \autoref{lemma:alpha_off_diagonal}, and we follow the notations set up in
\autoref{sec:extremal}. By \autoref{eqn:A_to_c}, if we set
\begin{equation*}
	A_{k+1} = 1, \qand A_i = 0, i \in [0,\floor{n/2}], i \neq k+1,
\end{equation*}
we have that
$$
c_{k+2} = \alpha_{k+2,k+1}.
$$
It suffices to compute $c_{k+2}$ in the following.
Substituting this assigment of $A_i, 0 \le i \le \floor{n/2}$ into
\autoref{eqn:Gleason_A} and setting $x = 1$, we have that
$$
y^{k+1} + \sum_{j = \floor{n/2}+1}^{n}A_jy^j =
\sum_{i=0}^{\floor{n/2}}c_i(1+(q-1)y)^{n-2i}(y(1-y))^i.
$$
Dividing by $(1+(q-1)y)^n$, we have that
$$
\frac{y^{k+1}}{(1+(q-1)y)^n} =
\sum_{i=0}^{\floor{n/2}}c_i\qty(\frac{y(1-y)}{(1+(q-1)y)^2})^i -
\frac{1}{(1+(q-1)y)^n}\sum_{j=\floor{n/2}+1}^n A_jy^j.
$$
Let $g(y) = \frac{y(1-y)}{(1+(q-1)y)^2}$. It is easy to check that $g(0) = 0$
and $g^\prime(0) \neq 0$. Now we may apply \autoref{lemma:LagrangeBurmann}.
Let $L(y) = \frac{1}{(1+(q-1)y)^n}\sum_{j=\floor{n/2}+1}^nA_jy^j$.
Note that
$$
L(y) = \sum_{j \ge \floor{n/2}+1}u_jy^j,
$$
where $u_j, j \ge \floor{n/2}+1$ are some coefficients whose value we do not
care.
Thus, the expansion of $L(y)$ in terms of $g(y)$ must be of the form
$$
L(y) = \sum_{j \ge \floor{n/2}+1}v_jg(y)^j,
$$
where $v_j, j \ge \floor{n/2}+1$ again are some coefficients whose value we do
not care.
In conclusion, the expansion of $\frac{y^{k+1}}{(1+(q-1)y)^n}$ in terms of $g(y)$
is of the form
$$
\frac{y^{k+1}}{(1+(q-1)y)^n} = \sum_{i=0}^{\floor{n/2}}c_ig(y)^i - \sum_{j \ge \floor{n/2}+1}v_jg(y)^j.
$$
By \autoref{lemma:LagrangeBurmann},
\begin{align*}
	c_{k+2} &= \frac{1}{k+2}\qty[\qty(\frac{y^{k+1}}{(1+(q-1)y)^n})^\prime\qty(\frac{(1+(q-1)y)^2}{1-y})^{k+2}][k+1]
	 \\
	&= \frac{k+1}{k+2}\qty[\frac{y^{k}}{(1+(q-1)y)^{n-2k-4}(1-y)^{k+2}}][k+1] \\
	&\qquad -\frac{n(q-1)}{k+2}\qty[
	\frac{y^{k+1}}{(1+(q-1)y)^{n-2k-3}(1-y)^{k+2}}
	][k+1] \\
	&=
	\frac{k+1}{k+2}\qty(\frac{1}{(1+(q-1)y)^{n-2k-4}(1-y)^{k+2}})^\prime(0) -
	\frac{n(q-1)}{k+2} \\
	&= \frac{k+1}{k+2}((n-2k-4)(1-q)+k+2) + \frac{n(1-q)}{k+2} \\
	&= (2q-1)(k+1) + (1-q)n.
\end{align*}
The proof is completed as desired.

\bibliographystyle{IEEEtran}
\bibliography{bib.bib}			
\end{document}